\documentstyle[12pt]{article}

\begin{document}

\title{Conformally Invariant  Braneworld and the Cosmological Constant}

\author{E. I. Guendelman \\
Department of Physics, Ben Gurion University, Beer Sheva, Israel 84105}

\maketitle

\bigskip

\begin{abstract}

A six dimensional braneworld scenario based on a model describing the 
interaction of gravity, gauge fields and $3+1$ branes in a conformally 
invariant way is described. The action of the model is defined using a
measure of integration built of degrees of freedom independent of the
metric. There is no  need to fine tune any bulk cosmological constant or
the tension of the two (in the scenario described here) parallel 
branes to obtain zero cosmological constant,
the only solutions are those with zero 4-D cosmological constant. The two
extra dimensions are compactified in a "football" fashion and the branes
lie on the two opposite poles of the compact "football-shaped" sphere.

\end{abstract}

\section{The Model}

Recently there have been a great number of studies of the possibility that 
our Universe is built of one or more 3+1 branes and a higher dimensional bulk
component $^{1,2,3,4,5}$ . Standard model fields would be confined to the 
branes while gravity can propagate into the bulk also. In this paper, the
question of the cosmological constant will be addressed in a model of this 
kind.

One approach to the cosmological constant problem has been to exploit the 
possibility of using in the action a measure of integration independent 
of the metric $G_{AB}$, that is, different from
$\sqrt{-G}d^{D}x$ , where $G = det(G_{AB})$. See Refs. 6, 7, 8 .

Indeed instead of $\sqrt{-G}d^{D}x$ one can use $\Phi d^{D}x$, where 
\begin{equation}
\Phi = 
\epsilon^{A_{1}..A_{D}} \epsilon_{a_{1}..a_{D}} \partial_{A_{1}} 
\varphi _{a_{1}} ...\partial_{A_{D}} \varphi_{a_{D}}
\end{equation}

in $D$ - dimensional space-time. Here the $ \varphi _{a}$ fields 
are scalars and are treated as 
indendent degrees of freedom. Since $\Phi$ has the same transformation 
properties as $\sqrt{-G}$, it follows that also $\Phi d^{D}x$ is a scalar 
as well as $\sqrt{-G}d^{D}x$.

Then we can writte for example an invariant action of the form
\begin{equation}
S = \int{L} \Phi d^{D}x
\end{equation}

where $L$ is a scalar.
One can explore the possibility of using also $\sqrt{-G}d^{D}x$ in another
piece of the action, this "two measures" possibility will not be used here.

Notice that $\Phi$ is a total derivative, indeed we can writte

\begin{equation} 
\Phi =\partial _{A_{1}} ( \epsilon^{A_{1}..A_{D}} \epsilon_{a_{1}..a_{D}}
\varphi_{a_{1}}...\partial_{A_{D}} \varphi_{a_{D}})
\end{equation}

Therefore the shift 
\begin{equation}
 L \rightarrow  L + constant 
\end{equation} 

does not change the equations of motion.

Let us now describe what will be our choice for $L$: it should describe 
gauge fields, gravity and $3+1$ branes.

For the (bulk) gauge field contribution to $L$ we will consider 
a "square root" form $ s \sqrt{|F_{CD}F^{CD}|}$ where $F_{CD} = 
\partial_{C}A_{D} -  \partial_{D}A_{C}$ . This is a very interesting 
possibility: Such "square root gauge theory" has been shown to give rise
to string solutions $^{9}$  and therefore provides a theory which includes
strings as special types of excitations, while containing other types of 
excitations as well. The square root gauge theory was considered also as a 
model for confinement$^{8, 10}$.

When including gravity, the bulk dynamics will be then governed by

\begin{equation}
S_{bulk} = \int{L_{bulk}} \Phi d^{6}x , 
L_{bulk} = -\frac{1}{\kappa} R +  s \sqrt{|F_{CD}F^{CD}|}
\end{equation}

where

\begin{equation}
R = G^{AB} R_{AB},  R_{AB} = R^{C}_{ABC}
\end{equation}
 
and the curvature tensor is defined in terms of the connection 
coefficients through the relation

\begin{equation} 
R^{A}_{BCD} = \Gamma^{A}_{BC,D} - \Gamma^{A}_{BD,C} + 
\Gamma^{A}_{ED} \Gamma^{E}_{BC} - \Gamma^{A}_{EC} \Gamma^{E}_{BD}   
\end{equation} 

The relation between $\Gamma^{A}_{BC}$ and $G_{AB}$ is determined 
by the variational principle. s is a number.

Model (5) was studied in Ref. (8) and shown to provide a compactification
mechanism without the need of fine tuning a bulk cosmological constant 
term. The compactification takes place when $F_{AB}$ takes a monopole
expectation value in two extra dimensions.

Under the change 

\begin{equation}
\varphi_{a} \rightarrow  \varphi^{'}_{a} (\varphi_{b})
\end{equation}

which means that 
$\Phi \rightarrow$ 
$ |\frac{\partial \varphi^{'}_{a}}{\partial \varphi_{b}}|\Phi$ we 
can achieve invariance 
of the action defined by (2), (5), (6), (7) if $G^{AB}$ is 
also transformed according to

\begin{equation} 
  G^{AB} \rightarrow 
 |\frac{\partial \varphi_{a}}{\partial \varphi^{'}_{b}}|G^{AB} 
\end{equation}

In this case we can recover the General Relativity form if by means of the
conformal invariance displayed before, we choose the gauge
$\Phi$$ =$$\sqrt{-G}$ . 

Notice that the conformal invariance (8), (9) is possible because both terms 
in $L$ in (5) have the same homogeneity in $G^{AB}$. They are both
homogeneous of degree one in this variable, so their transformation can be
simultaneously compensated by the transformation of the  measure $\Phi$.

To construct a brane scenario we must of course study the introduction of a
brane term to $L$. We will see that the action of a $3+1$ brane embedded in a
six dimensional space is consistent with the conformal symmetry (8)- (9).

Before considering the action of a $3+1$ brane in the context of an action
of the form (1) and (2), let us review what is the action in the context
of the standard formulation. In this case

\begin{equation}
S^{standard}_{4} = \int d^{4}\sigma \sqrt{-g} l_{4} = 
\int d^{6}x \sqrt{-G} L_{4}
\end{equation}

here $g = det (g_{\mu \nu})$ and $g_{\mu \nu}$ being the metric pulled 
back to the brane world volume.

\begin{equation}
g_{\mu \nu} = G_{AB} x^{A}_{,\mu} x^{B}_{,\nu} 
\end{equation}  

and

\begin{equation}
 L_{4} = \int d^{4}\sigma \frac {\sqrt{-g}}{\sqrt{-G}}
\delta^{(D)} (x - x(\sigma)) l_{4}
\end{equation}

and we will consider the case where $D=6$ and  $l_{4} = T = constant$ . This
constant $T$ has the interpretation of '$3+1$ surface tension'.

We can incorporate (12) into the general form (1), (2)  by 
changing the measure of integration:
 $d^{6}x\sqrt{-G} \rightarrow d^{6}x\Phi$ , so that the brane contribution 
will be now 

\begin{equation}
S_{4} = \int d^{6}x \Phi  L_{4}
\end{equation}

where $L_{4} $ is given by (12) for $l_{4} = T = constant$ .

It is very important to see the very special role of the dimensionalities
$3+1$  of the brane and $6=5+1$ of the embedding space: indeed  
$L_{4} $ as defined by (12) is of degree one in $G^{AB}$ only for the very
special choice $D=6$.
So that 
\begin{equation}  
S = S_{bulk} + S_{4} 
\end{equation}

with $S_{bulk}$ given by (5) and $S_{4}$  given by (12), (13), 
has the symmetry 
(8), (9). Notice that by using the symmetry (8), (9) we can set the gauge
$\Phi = \sqrt{-G}$ and in this case $S_{4} =S^{standard}_{4} $. This once
again, only for a $3+1$ brane when embedded in a six dimensional space.

\section{A zero four dimensional cosmological constant without fine tuning}

The fact that $L = L_{bulk} + L_{4} $, where 
 $L_{bulk}$ and $ L_{4} $ are given by (5) and (12), is homogeneous in  
$G^{AB}$ with homogeneity one
implies,

\begin{equation}
G^{AB}\frac{\partial L}{\partial G^{AB}} = L
\end{equation}  

(and a similar equation holds also separately for $L_{bulk}$ and $ L_{4} $)
The variation of the action with respect to the $\varphi_{a}$ fields
gives the equation

\begin{equation}
A^{A}_{a} \partial_{A} L = 0
\end{equation} 

where 
\begin{equation}
A^{A}_{a} = \epsilon^{A A_{2}..A_{6}} \epsilon_{a a_{2}..a_{6}} 
\partial_{A_{2}} \varphi _{a_{2}} ...\partial_{A_{6}} \varphi_{a_{6}}  
\end{equation}

since one can  easily see that  
$det (A^{A}_{a}) = \frac{6^{-6}}{6!} \Phi ^{6}$,
we have that if $\Phi \neq 0$, then $\partial_{A} L = 0 $,
which means that

\begin{equation}
L = M = constant
\end{equation} 

Taking a variation of the action (2) with respect to a conformal transformation, 
i.e. to a transformation of the form $G^{AB} \rightarrow \Omega^{2}(x)G^{AB}$,
we obtain, for the case that $L$ is an homogeneous function of $G^{AB}$
(of non trivial homogeneity),
that $L = 0$, so that the constant $M$ in (18) equals zero.

The constant of integration $M$, if it were different from zero would have
spontaneously broken the conformal invariance, since $L$ changes under a
conformal transformation (as mentioned  $L$ is of homogeneity one in
$G^{AB}$ ) and $M$ does not (is fixed by the boundary conditions).
If we work with theories with  global scale invariance, there
is the possibility of spontaneous breaking of global scale invariance by
the appearence of non zero constant of integration $M$. See Refs.
(7)

Separating Gravity and matter pieces of the action, we define

\begin{equation}
L =  -\frac{1}{\kappa} R + L_{matter}
\end{equation}

 and we obtain, from the variation with respect to $G^{AB}$ (using the fact
that the fields in the measure $\Phi$ are independent of $G^{AB}$),

\begin{equation}
R_{AB} = \kappa \frac{\partial L_{matter}}{\partial G^{AB}}
\end{equation}

Alternatively, one can perform the variation with respect to $G^{AB}$
after setting the "Einstein" gauge $\Phi = \sqrt{-G}$. In this case 
one obtains the Einstein's equations corresponding to the matter lagrangian
$L_{matter}$. These equations are identical to (20) if use is made of the 
fact that $L_{matter}$ is homogeneous of degree one in  $G^{AB}$.

In any case, even before we start to solve in detail the equations of
motion, we see something remarkable from eq. (20): Let us consider a product
metric of the form

\begin{equation}
ds^{2} = g_{\mu \nu} (x^{\alpha}) dx^{\mu} dx^{\nu} + 
\gamma_{ij} (x^{k}) dx^{i} dx^{j} 
\end{equation}

where the ordinary dimensions are labeled with greek indices, 
$\mu , \nu , \alpha = 0, 1, 2, 3 $ and the extra dimensions with small 
latin indices,  $ i, j, k  = 4, 5 $ 
and consider the case when $F_{AB}$ takes non zero values only in the extra 
dimensions  $ i, j, k = 4, 5 $  and let us consider the branes to be oriented
in the $3+1$ hyperplanes orthogonal to those of the two extra dimensions
, i.e. hyperplanes  $ x^{i} = constant $.

In this case we see that in $ s \sqrt{|F_{CD}F^{CD}|} $ only the extra 
dimensional metric appears and like wise the same is true for
$S_{4}$, because there, in the ratio $\frac{g}{G}$  the four dimensional
metric is cancelled.

Therefore (20) implies that both $ s \sqrt{|F_{CD}F^{CD}|} $ and $S_{4}$
curve only the extra dimensions. A solution containing four dimensional
flat Minkowskii space is possible without fine tuning whatsoever!,
i.e. we can take

\begin{equation} 
ds^{2} = -dt^{2} +dx^{2} + dy^{2} + dz^{2} + \gamma_{ij} (x^{k}) dx^{i} dx^{j}
\end{equation}  

\section{The explicit braneworld solutions}

Given the metric (22), choosing the Einstein form $\Phi = \sqrt{-G}$, we 
can see that, if we want an extra dimensional space of constant curvature
(at the points where there are no branes), we must consider a field strength
which has expectation value in the extra dimensions of the form, as has
been done in many previous studies of spontaneous compactification$^{11}$
\begin{equation}
F_{ij} = B_{0} \sqrt{\gamma} \epsilon_{ij}
\end{equation}

where $B_{0} $ is a constant.  
This automatically gives $F^{AB}F_{AB} =  2B^{2}_{0} = constant$ and from
(20) and (23) a constant extradimensional curvature appears.

This configuration satisfies also the field equations for the gauge fields

\begin{equation}
\partial_{A}(\sqrt{-G} \frac{F^{AB}}{\sqrt{|F^{CD}F_{CD}}|} ) = 0 
\end{equation}

If there are no branes ($T = 0$, $S_{4} = 0$), we take the extra dimensions
to have a spherical shape,

\begin{equation} 
\gamma_{ij} (x^{k}) dx^{i} dx^{j} 
= b^{2} (d\theta^{2} + sin ^{2} \theta d\phi ^{2}), b = const.
\end{equation}  

and then eq. (20) implies that there is the following relation:

\begin{equation}
   b^{2}    = \frac{\sqrt{2}}{s \kappa B_{0}}   
\end{equation}

When branes are present, we can keep still a gauge field configuration
satisfying (23), but then the extra dimensional part of the metric has to be
changed, to a "foot balllike configuration" in the language of Carroll and 
Guica$ ^{12}$, see also Ref. 13.

We consider two branes located at opposite poles of the spherical extra 
dimensions. Following the analysis of Carroll and Guica $^{12}$, 
we represent the extra dimensions as
\begin{equation}
\gamma_{ij} (x^{k}) dx^{i} dx^{j} = \psi (r) (dr^{2} + r^{2} d\phi^{2})
\end{equation}

To describe brane sources we need to introduce singularities at the north 
pole $r = 0$ and at the south pole ($ r = \infty$, but a new coordinate
system should be used there). The standard two dimensional delta 
function with respect to integration measure  
$rdr d\phi$ is

\begin{equation}
\delta^{(2)} (r) = \frac{1}{2 \pi} \nabla ^{2} ln r
\end{equation}

where $\nabla ^{2} f = f{''} + \frac{1}{r} f{'} $, $'$ being derivation with 
respect to $r$.

For the metric defined by equations(22) and (27) we have,

\begin{equation}
R_{\mu \nu} = 0, R_{rr} = -\frac{1}{2} \nabla ^{2} \psi,
 R_{\phi \phi} = -\frac{r^{2}}{2} \nabla ^{2} \psi,  
R = -\frac{1}{\psi} \nabla ^{2} \psi  
\end{equation}

From (20) and from the fact that 
$G^{AB} \frac{ \partial L_{matter}}{ \partial G^{AB}}$$ =$$ L_{matter}$,
we get that

\begin{equation}
R = \kappa L_{matter}
\end{equation}

equation that gives (using the representation (28) for the two dimensional
delta function and eqs. (5) and (12) for the matter lagrangians of the gauge
fields and $3+1$ brane)

\begin{equation}
-\frac{1}{\psi} \nabla ^{2} \psi  = s \kappa \sqrt {2} B_{0}
+ \frac{ \kappa T }{ 2 \pi \psi} \nabla ^{2} (ln r)    
\end{equation} 

This equation, which appears also in the context of
$2+1$ dimensional gravity,  has the solution$^{14}$ (the different constants
in (31) having a different meaning in the corresponding eq. in Ref. 14 
of course)

\begin{equation}
 \psi = \frac{4 \alpha ^{2} b^{2}}{ r^{2} [( \frac{r}{r_{0}})^{\alpha} + 
( \frac{r}{r_{0}}) ^{-\alpha} ] ^{2}} 
\end{equation} 
where  $r_{0}$ is an arbitrary parameter and
\begin{equation}
 \alpha = 1 - \frac{\kappa T}{4 \pi} , 
b^{2} = \frac{\sqrt{2}}{ s \kappa B_{0}}  
\end{equation} 

Such a metric can be transformed into the form (where $r_{0}$ goes away),

\begin{equation}
\gamma_{ij} (x^{k}) dx^{i} dx^{j}
= b^{2} (d\theta^{2} + \alpha ^{2} sin ^{2} \theta d\phi ^{2}) 
\end{equation}

where $\phi$  ranges from $0 $ to $ 2 \pi $, or equivalently
\begin{equation}
\gamma_{ij} (x^{k}) dx^{i} dx^{j}    
= b^{2} (d\theta^{2} + sin ^{2} \theta d\bar{\phi} ^{2})
\end{equation} 

where $\bar{\phi}$ now ranges from $0 $ to $ 2 \pi \alpha < 2 \pi $.
The effect of the branes (at the opposite poles of the extra dimensional
sphere and having the same tension)
is just to produce a deficit angle and changing the shape of the
extra dimensional into a football like sphere.

It should be pointed out that the solutions with
 $F^{AB}F_{AB} = constant$ (used here) are not the most general solutions.
In four dimensions the square root gauge theory allows string like
solutions $^{9}$ . The generalizations of the string solutions in the 
square root gauge theory in six dimensions are $3+1$ brane solutions, as
we will see elsewhere, so that in fact the  $3+1$ branes do not have to be 
added to $S_{bulk}$ defined in eq.(5).

A complete study of the solutions of the square root gauge theory plus
gravity will be done elswhere, but still, as we have seen in section 2, 
which does not depend on the details of the solutions, we have that as long
as $F_{AB}$ takes expectation values in the extra dimensions only, does
not matter in which precise way, only the extra dimensions curve and the
four dimensional space remains flat.

\section{Discussion and Conclusions}

We have seen that it is possible to construct a totally conformally
invariant six dimensional brane world scenario by means of the introduction 
of a new measure of integration in the action. This new measure of integration
depends of degrees of freedom independent of the metric. The model includes
gravity, gauge fields and $3+1$ branes.

We have also seen that if the gauge fields take expectation values in
the extra dimensions and if the $3+1$ branes are orthogonal to the
extra dimensions, we get that the four dimensional subspace can remain
flat without need of any fine tuning, only the extra dimensions are curved. 
For this it is essential that $\sqrt{-G}$ did not enter in the measure of
integration of the action ($\Phi = \sqrt{-G}$ can be choosen as a particular
"gauge" however).

An explicit solution with the "football " 
 like compactification for the extra 
dimensions $^{12, 13}$  and flat four dimensional space which exemplifies
the above has been displayed. More general solutions will be studied 
elsewhere.

Of course to describe the real Universe one has to go beyond this model
and introduce spontaneous symmetry breaking of scale invariance, a small
vacuum energy to describe the presently accelerated Universe, as has been
done in the context of alternative measure theories in four dimensions $^{7}$.
This six dimensional model represents nevertheless significant progress,
since it allows us to see the applicability of the "modified measure"  
approach to braneworld scenarios and therefore for a starting point of a
new direction in the research of these kind of theories.


\begin{thebibliography}{1}

\bibitem{1} V.A. Rubakov and M.E. Shaposhnikov, Phys. Lett. B125, 136 (1983);
M. Visser, Phys. Lett. B 159, 22 (1985).
%
\bibitem{2} P.Horava and E.Witten, Nucl. Phys. B460, 506 (1996); 
 P.Horava and E.Witten, Nucl. Phys. B475, 94 (1996). 
%
\bibitem{3} N.Arkani-Hamed, S.Dimopolos and G.Dvali, Phys. Lett. B 429, 263
(1998); I. Antoniadis,  N.Arkani-Hamed, S.Dimopolos and G.Dvali,
Phys. Lett. B 436, 257 (1998);  N.Arkani-Hamed, S.Dimopolos and 
J. March-Russell, Phys. Rev. D63, 064020 (2001).
 
%
\bibitem{4} L.J. Randall and R. Sundrum, Phys. Rev. Lett. 83, 3370 (1999);
L.Randall and R.Sundrum,  Phys. Rev. Lett. 83, 4690 (1999).
%
\bibitem{5} V.A. Rubakov, hep-ph/0104152. 
%
\bibitem{6} For a review on issues not connected to scale invariance
in this kind of theories see E.I.Guendelman and A.B.Kaganovich,
Phys. Rev. D60: 065004 (1999).
%
\bibitem{7}  E.I.Guendelman Mod. Phys. Lett. A14, 1043 (1999). Applications 
the to fermion family problem were done in E.I.Guendelman and A.B.Kaganovich,
 Int. Journ. Mod. Phys. A17, 417 (2002); E.I.Guendelman and A.B.Kaganovich,
Mod. Phys. Lett. A17, 1227 (2002) and for further ideas on cosmology and full
list of references see E.I.Guendelman and O.Katz, gr-qc/0211095. 
%
\bibitem{8} E.I.Guendelman, Phys. Lett. B412, 42 (1997)
%
\bibitem{9} A. Aurilia, A. Smailagic and E.Spallucci, Phys. Rev. D47, 2536
(1993) and references
%
\bibitem{10} N. Amer and E.I.Guendelman,  Int. Journ. Mod. Phys. A15, 4407
(2000).
%
\bibitem{11} See for example, E.Cremmer and J.Sherk, Nucl. Phys. 
B108, 409 (1976);Z.Horvah, L.Palla, 
E. Cremmer and J.Sherk, Nucl. Phys. B127, 57 (1977);
P.G. Freund and M.A. Rubin, Phys. Lett B97, 233 (1980); 
S.Ranjdar-Daemi, A.Salam and J. Strathdee, Nucl. Phys. B214, 491 (1983). 
%
\bibitem{12} S.M.Carroll and M.Guica, hep-th/0302067  
%
\bibitem{13} I.Navarro, hep-th/0302129.  
% 
\bibitem{14} S. Deser  and R.Jackiw, Annals Phys. 153, 405 (1984).  
% 
\end{thebibliography}
\end{document}